\documentstyle[12pt]{article}

\newcommand{\mib}[1]{\mbox{\boldmath$#1$}}

\title{Proton Elastic Scattering \\
       and Neutron Distribution of Unstable Nuclei 
       }
\author{K.Kaki \\
Department of Physics, Shizuoka University, Shizuoka 422-8529, Japan \\
        \normalsize
        tel:+81-54-238-4744, fax:+81-54-238-0993 \\
        \normalsize
        e-mail:kkaki@sci.shizuoka.ac.jp \\ \\
        S.Hirenzaki \\
Department of Physics, Nara Women's University, Nara 630-8506, Japan \\
        }
\date{}
\begin{document}
\titlepage
\maketitle

\begin{abstract}
We study theoretically how we can determine the neutron density 
distributions of 
unstable nuclei from proton elastic scattering. We apply the relativistic 
impulse model to 
study the sensitivities of the observables to the density distributions
which are expressed in Woods-Saxon form.    
We find that the both radius and diffuseness of densities can be determined from 
restricted elastic scattering data in principle.  We think this result is helpful 
to design future experiments.  
\end{abstract}
\newpage

\baselineskip=24pt

\section{Introduction}

Since the successful use of the radioactive ion beam, the properties of 
nuclei far from stability line have been studied extensively in both 
theoretical and experimental ways \cite{1,2}.  So far clear evidences of neutron 
skin/halo structure are found for $^{11}$Li \cite{3} and sodium isotopes \cite{4}. 
There are 
also many unstable nuclei which are expected to have interesting structures. 
Thus, to study unstable nuclei is one of the most exciting subjects of 
nuclear physics.  

In experiments, radioactive ions are usually provided as secondary beam 
and used for nuclear reactions.  Since the targets should be heavy enough to 
use the beam energy efficiently, it is difficult to use electron 
scatterings as a tool to investigate the nuclear structure.  Furthermore, 
intensities of the secondary beams are much lower than the primary beams 
used to study the stable nucleus.  Because of these facts we think that 
we need a kind of prescriptions to determine the densities of unstable 
nucleus from the very restricted experimental data.  

So far, to study the structure of the unstable nuclei we have used several 
kinds of data such as 
interaction cross sections, neutron-removal cross sections, 
transverse momentum distributions of projectile fragments, and so on \cite{1,2}. 
Using theoretical models of the nuclear reactions, radii of some of unstable 
nuclei are extracted from the data.  Even we can extract the radii of both proton
and neutron distributions
when we have the information 
of charge distribution \cite{4,5}.  However, still it seems difficult to determine 
shapes of density distributions from experimental data.  

In this paper, we try to develop a prescription to determine both radius 
and diffuseness of densities from restricted information of proton elastic 
scattering.  Proton elastic scattering on unstable nuclei has been performed 
several times \cite{6,7}.  Since theoretical investigations of the data indicate 
usefulness of the proton elastic scattering \cite{8}, it seems very interesting 
to theoretically study the sensitivities of observables to nuclear structure 
and to propose which should be observed to determine both radius and 
diffuseness.  We expect 
that this result is much helpful to design the future experiments.  

In section 2, we describe relativistic impulse model which we have used 
to make a relation between nuclear density distributions and Dirac optical 
potentials.  We show numerical results and investigate the 
sensitivities of the observables to nuclear density parameters in section 3.  
We summarize this paper in section 4.

\section{Formulation}

We calculate Dirac optical potentials using nuclear density distributions
based on the relativistic impulse approximation (RIA) .
Observables of proton elastic scatterings are obtained by solving the 
Dirac equation with the RIA optical potentials.  This model is known to 
describe the proton-nucleus elastic scattering sufficiently 
well for wide range 
of stable-nucleus targets at intermediate energies \cite{9}.

The Dirac equation with an RIA optical potential term is given as; 
\begin{eqnarray}
\left\{ E \gamma^0 - \mib{\gamma} \cdot \mib{p} - M \right\}
\psi ( \mib{p} ) = \hat U( \mib{p}, \mib{p}' ) \psi ( \mib{p}' ),
\end{eqnarray}
where $E$ is the energy of proton in center of mass frame, $M$ is the nucleon 
mass, and $\hat U( \mib{p},\mib{p}' )$ is the RIA optical potential 
which is written as follows; 
\begin{eqnarray}
\hat U( \mib{p}, \mib{p}' ) = \int {d^3 k \over (2\pi)^3}
  \sum_{\alpha} \bar \psi_{\alpha} ( \mib{k}+{1 \over 2}\mib{q} ) 
\ ( - \hat M ) \     \psi_{\alpha} ( \mib{k}-{1 \over 2}\mib{q} ).
\end{eqnarray}
Here, $\psi_{\alpha}$ indicates the relativistic nucleon wave function in the 
target, $\hat M$ is Feynman nucleon-nucleon amplitudes, and 
$\mib{q} = \mib{p}-\mib{p}'$ is the momentum transfer.  

We use the optimal factorization and rewrite the optical potential to 
so-called 't$\rho$' form; 
\begin{eqnarray}
\hat U( \mib{p'}, \mib{p} )  =  {1 \over 4} {\rm Tr}_2 ( - \hat M )
                                  \hat \rho( \mib{q} ), 
\end{eqnarray}
where the Tr$_2$ indicates to take summation of spin of nucleons inside 
the target.  Nuclear density in eq. (3) is written as; 
\begin{eqnarray}
\hat \rho( \mib{q} ) & = & \rho_S(q) 
                     + \gamma_2^0 \rho_V(q) 
                     -{\mib{\alpha}_2 \cdot \mib{q} \over 2M} \rho_T(q).
\end{eqnarray}
In numerical calculation, we take model densities in coordinate 
space and calculate each type of $\rho(\mib{q})$ in eq. (4) 
by Fourier transformation.  

The Feynman amplitudes in eqs. (2) and (3) are described as an expansion 
by complete set of Lorentz covariants as; 
\begin{eqnarray}
\hat M (p_1,p_2 \rightarrow p_1',p_2' )  = 
      \sum_{\rho_1, \rho_2, \rho_1', \rho_2'} 
      \Lambda^{\rho_1'}(p_1') \Lambda^{\rho_2'}(p_2')
      \sum_{n=1}^{13} M_n^{\rho_1 \rho_2 \rho_1' \rho_2'}
                         ( p_1,p_2 \rightarrow p_1',p_2' )
      \kappa_n \Lambda^{\rho_1}(p_1) \Lambda^{\rho_2}(p_2),
\end{eqnarray}
where $\Lambda^{\rho}(p)$ is the covariant projection operator defined as  
$\Lambda^{\rho}(p) = {1 \over 2M}( \rho \ \gamma^{\mu} p_{\mu}+M)$. 
The $\rho$ indicates + or - which distinguishes
positive- and negative-energy components of the initial and final 
states.   The $\kappa_n$ is unsymmetrized Fermi covariants shown in Table 1. 
In the present calculation, we use the IA2 parameterization of the 
amplitudes in ref. \cite{10}.

\section{Numerical Results} 

In this section, we show numerical results obtained by the theoretical 
model described in the last section.  First, we have calculated the 
observables of the proton elastic scattering of $^{40}$Ca at T=300MeV 
and compared them with data in order to 
check our framework.  We calculate the RIA optical 
potential using the density distribution of relativistic mean field theory 
(RMFT) \cite{11}.  The results are shown in Fig.1.  
As already discussed \cite{12}, 
we see that the present framework ( the first-order RIA ) gives reasonably good
predictions for both differential cross section and analyzing power.

We, then, study the sensitivities of the observables to the shape of neutron 
density distribution using the model densities.  We use Woods-Saxon form 
defined as; 
\begin{eqnarray}
\rho(r) = {\rho_0 \over {1 + e^{(r-r_0)/a}}},
\end{eqnarray} 
where $r_0$ is a radius parameter and $a$ is a diffuseness parameter.  As an 
example, we consider extremely neutron-rich nuclei $^{60}$Ca and calculate the 
observables.  The $r_0$ parameter is fixed to be 4.48 (fm) which is the half 
density radius of RMFT result and the $a$ parameter is varied from 0.5 to 2.0 
(fm). 
We have used the same $r_0$ and $a$ parameters for both the scalar 
and the vector densities appeared in eq. (4).
We do not include the tensor density in this model calculation using 
Woods-Saxon densities since its contribution to the cross section is 
known to be small.  We use the RMFT proton densities here.  We show in Table.2 
the calculated root-mean square radius and reaction cross section of each density. 
The differential cross sections and spin observables are shown in Fig.2.   

As shown in Table.2, we find that the reaction cross section is very 
sensitive to the root-mean-square radius as we expected.  
The cross section increases monotonically when we increase the diffuseness
parameter and namely the root-mean-square radius of the nucleus.
We expect to 
obtain important information on nuclear size by observing reaction cross section.  
In Fig. 2, we show the angular distribution of observables for each 
diffuseness value.  We can see interesting behavior of the differential 
cross section.  'The differential cross section has a dip in smaller 
angles for smaller diffuseness parameters, namely for smaller 
root-mean-square radii.' 
We usually think that the dip spacing in the differential cross section can
be estimated using a relation $\Delta q \cdot R \sim \pi$, where $\Delta q$
is the difference of momentum transfer corresponding to difference of the
scattering angle and R is the nuclear radius.  Thus, we expect the first
dip angle of the differential cross section will be more forward for
larger nucleus.  The calculated results shown in Fig. 2 have opposite
tendency to this naive expectation, which seems to be useful to
determine both radius and diffuseness parameter from data.

In order to clarify the validity of the two observables
mentioned above, the reaction cross sections and the first dip angles in
the differential cross sections, we calculated the both quantities for
various neutron densities for the typical stable nucleus $^{40}$Ca.  We
have assumed the Woods-Saxon form for neutron densities and varied the
parameters in the range of $3.2 \le r_0 \le 4.3$ [fm] and $0.1 \le a
\le 0.8$ [fm].  
We use the proton density determined by the relativistic mean filed theory
(RMFT) which has been well established \cite{13}.

We show the results in Fig. 3.  Each point corresponds to a result calculated 
with a different neutron density. 
 First, we can see clearly that there exist strong
correlation between root-mean-square radius $r_{rms}$ and reaction cross
section $\sigma_{R}$ as expected.  This correlation is almost independent
to the Woods-Saxon density parameters as can be seen in the Fig.3 and
thus, is expected to be independent to detail structures of the nucleus.
This fact indicates that we can determine $r_{rms}$ by reaction cross
sections in model independent way.

On the other hand, in the same Fig.3, we see that the first dip angles 
of differential cross sections
depend on $r_{rms}$ in much different way from $\sigma_{R}$. The first dip
angle varies in certain range for a fixed $r_{rms}$.  Thus, in principle,
we can determine
the shape of the neutron density distribution, both
radius and diffuseness,
using the two experimental data, reaction cross
section and first dip angle of differential cross section.  We can fix
the $r_{rms}$ from the reaction cross section and, then, fix the unique
combination of the radius and diffuseness parameters by knowing the first
dip angle of the differential cross section.

In order to check practical applicability of this idea, we show in Fig. 4
the contour plot of (a) the reaction cross sections and (b) the first dip angles
of the differential cross section in the $r_0$ (radius parameter) - $a$
(diffuseness parameter) plane of the Woods-Saxon neutron-density distribution.  
We consider $^{40}$Ca as a typical well-known stable nucleus to check how well
we can fix the neutron densities from the reaction cross section and the
first dip angle. 
The proton density determined by RMFT \cite{13} is used in this calculation.  
We find that the two
contour plots show much different behavior and we can find a unique point
in the $r_0$-$a$ plane using the two kinds of data.  In the figure, we 
show the experimental data point, $\sigma_R=51$ [fm$^2$] 
and $\theta_{dip}=15.3$ [deg],
by solid circle and experimental uncertainties by hatched area.  We have
assumed  2 \% errors for experimental data.  In the actual experiments of
proton elastic scattering of unstable nuclei, the experimental errors
estimated to be several percent for reaction cross sections and several
milliradian for the first dip angles in the case of $^{56}$Ca \cite{14}.

We show in Fig.5 neutron densities which are obtained from the reaction cross
sections and the first dip angles.  
For comparison, we also show the RMFT proton and neutron densities in the
same figure.  We find the neutron density shown by the thick solid line,
which reproduce both experimental value, agrees extremely well with the RMFT
neutron density around nuclear surface where we expect to find
interesting nuclear structure like neutron halo and skin.  They agree
reasonably well nuclear inside, too.  Then, in order to investigate the
effects of the errors included in the observations of the reaction cross
sections and the first dip angles, we show the neutron densities
determined by the  2 \% different values of these observables.
Difference of three solid lines indicate the ambiguity
of neutron density determination by the present method
in the case where we have 2 \% error in both data.

We find that we can determine the neutron density
distribution only using the reaction cross sections and the first dip
angles of proton elastic scattering if they can be measured accurately
enough.  This is very interesting since we have a possibility to fix both
radius and diffuseness parameter by only two experimental numbers which can
be measured practically.  We do not need to measure very small cross
sections at larger angles, which are necessary sometimes to perform
standard $\chi^2$ fit.

As an example of the neutron rich nucleus, we consider $^{60}$Ca and applied
the same procedure for $^{40}$Ca explained above. 
We have used the proton
density distribution calculated by RMFT \cite{11} for the $^{60}$Ca target.  
The reaction cross section and
the first dip angles are calculated using the RMFT neutron density first
and they are 75.43 [fm$^2$] and 12.22 [deg], which are used instead of
experimental values.  We then replace the neutron densities to the
Woods-Saxon form and draw the contour plot as shown in Fig. 6.  We
assumed the 2 \% errors for both the reaction cross sections and
the first dip angles, which are shown as hatched area.

In Fig. 7, we compare the obtained neutron densities of the Woods-Saxon
form with the RMFT neutron density used to calculate the cross section
and the dip angle. 
We find that their agreement is reasonably well.  
Although they differ each other about 10 \% at $r<3$ (fm), they show excellent
agreement again around nuclear surface.
By comparing the three solid lines, we can also see the ambiguity
of neutron density determination in case we have 2 \% experimental errors
in data of the reaction cross sections and the first dip angles of the
differential cross section.
This example 
seems to indicate that our prescription is very useful to determine the 
neutron density distribution of neutron rich nucleus from restricted data: 
reaction cross 
section and the first dip position of differential cross section.  

\section{Summary}

We have studied proton elastic scattering theoretically to find 
a suitable prescription to obtain the nuclear density distributions from restricted 
experimental data.  We think 
 that this prescription must be very useful to study 
structure of unstable nuclei, for which we can not apply electron scattering.  

We have used relativistic impulse model to calculate 
the observables which are known to describe the elastic scattering well.  
Sensitivities of observables to density shape are studied with the 
Woods-Saxon model densities.  
We have found that the reaction cross section and the first dip
position of the differential cross section depend on the neutron density
distribution in much different way.
We have shown 
that we can determine both radius and diffuseness parameters in terms of
the dip position and reaction cross section.
The root-mean-square radius is directly deduced from the reaction cross section,
and the whole density shape can be determined by help of the dip position.

In order to demonstrate applicability of the prescription, 
we have tried to determine the neutron density distribution of
$^{40}$Ca and $^{60}$Ca from the reaction cross section and the angle of the
first dip of the proton elastic scattering. 
We find that the density
distribution can be determine well if the both data can be observed
accurately enough.  We think that this 
prescription is very useful to design future experiments.  

In this paper, we focused our attention on the neutron density distribution
assuming that the proton densities are known precisely.  This
assumption is not always correct.  In the study of the unstable nuclei,
in many cases, the proton density is not known.  In such cases, however, we
can apply the prescription described in this paper to determine the shape
of the matter distribution.  This is also valuable since only
the root-mean-square radius of the matter has been determined in 
the previous studies. 

\section{Acknowledgement}

We would like to thank Dr. Y.Sugahara for providing a code 
of extended RMFT and to Prof. H.Toki for valuable discussions.
We are also grateful to Prof.R.Seki and Prof. I.Tanihata for their
helpful comments.
One of us 
(S.H.) acknowledges many discussions with Prof. K.Kume and
Dr. H.Fujita.

Numerical calculations in this paper have been performed using 
workstations at Information Processing Center of Shizuoka University.

\newpage
Table 1. Definition of $\kappa_n$ in eq. (5).
The momentum $Q_{ij}$ is defined as 
$Q_{ij}={1 \over 2m} (p'_i + p_j )$ where $p$ and $p'$
correspond to initial and final momentum, respectively.
$\tilde S$ is the Fierz exchange operator defined as
$\tilde S={1 \over 4}( \kappa_1 + \kappa_2 + \kappa_3 + \kappa_4 - \kappa_5 )$.
\vspace{0.5cm}

\begin{tabular}{crcl|crcl} \hline
\hspace{1cm} & $n$ & \hspace{1cm} & $\kappa_n$ \hspace{2cm} & 
\hspace{1cm} & $n$ & \hspace{1cm} & $\kappa_n$ \hspace{2cm} \\ \hline
& 1   & & $I$                           & & 6   & & $ \gamma_2 Q_{11}$ \\
& 2   & & $\gamma_1 \gamma_2$           & & 7   & & $ \gamma_1 Q_{22}$ \\
& 3   & & $\sigma_1 \sigma_2$           & & 8   & & $ P \gamma_2 Q_{11}$ \\
& 4   & & $\gamma_1^5 \gamma_2^5 = P $  & & 9   & & $ P \gamma_1 Q_{22}$ \\
& 5   & & $\gamma_1^5 \gamma_1 \gamma_2^5 \gamma_2$
                             & & 10  & & $\gamma_2 Q_{12} \tilde S$ \\
&     & &                    & & 11  & & $\gamma_2 Q_{21} \tilde S$ \\
&     & &                    & & 12  & & $P \gamma_2 Q_{12} \tilde S$ \\
&     & &                    & & 13  & & $P \gamma_1 Q_{21} \tilde S$ \\ \hline
\end{tabular}

\newpage
Table 2. Root-mean-square radii and reaction cross sections calculated 
using the Woods-Saxon neutron densities for $^{60}$Ca case. 
Results with RMFT density are also shown.
We use RMFT proton density \cite{11}
and fix the radius parameter of Wood-Saxon neutron
densities to $r_0=4.48$ (fm)  for all cases shown in this table. 
\vspace{0.5cm}

\begin{tabular}{c||c|llll|c}\hline
        & diffuseness & \multicolumn{4}{c|}{root-mean-square radius (fm)} 
                               & reaction cross section \\ \cline{3-6}
density & parameter   & \multicolumn{2}{c}{neutron}
                      & \multicolumn{2}{c|}{proton} &  (fm$^2$) \\ 
        & ( fm )      & \ scalar \ & \ vector \
                      & \ scalar \ & \ vector \ & \\
\hline 
RMFT &  -        & 4.150 & 4.133 & 3.555 & 3.550 & 75.34 \\
\hline 
W-S  & 0.5       & 3.937 & 3.937 &       &       & 72.77 \\
     & 1.0       & 5.083 & 5.083 &       &       & 87.60 \\
     & 1.5       & 6.490 & 6.490 &       &       & 100.9 \\
     & 2.0       & 7.773 & 7.773 &       &       & 109.0 \\
\hline
\end{tabular}

\newpage

\centerline{\bf Figure Captions}
\begin{description}
\item[Figure 1]
Observables of proton - $^{40}$Ca elastic scattering at 
T=300MeV.  We show (a) differential cross section, (b) analyzing power, 
and (c) spin-rotation parameter. 
Solid curves show the calculated results with nuclear density obtained by
relativistic mean field theory (RMFT) \cite{11}. 
Solid circles indicate experimental data taken from ref.[15].

\item[Figure 2]
Observables of proton - $^{60}$Ca elastic scattering at 
300MeV using the Woods-Saxon model densities for neutron distributions.
We show (a) differential cross section, (b) analyzing power, and (c) 
spin rotation parameter. Results with the 
diffuseness parameter a=0.5, 1.0, 1.5, 2.0 (fm) are shown by solid line, 
dotted line,  dashed line, dash-dotted line, respectively.  
The RMFT density \cite{11} is used for proton distribution.

\item[Figure 3] 
Calculated reaction cross sections (solid triangles) and first dip angles
of the differential cross sections (solid circles) are shown as a function
of the
root mean square radius of neutron distribution
for proton-$^{40}$Ca elastic scattering at T= 300 MeV.  Proton density
distribution determined by RMFT is used \cite{13}. 
Neutron distribution is assumed to have Woods-Saxon
form which parameter is varied in the range of 3.2 $\le r_0 \le 4.3$ [fm]
and $0.1 \le a \le 0.8$ [fm].  Each point corresponds to the calculated
result with different neutron density distribution.  For guiding eyes,
we connect solid circles calculated with the same diffuseness parameters
by solid lines and those with the same radius parameters by dashed lines.

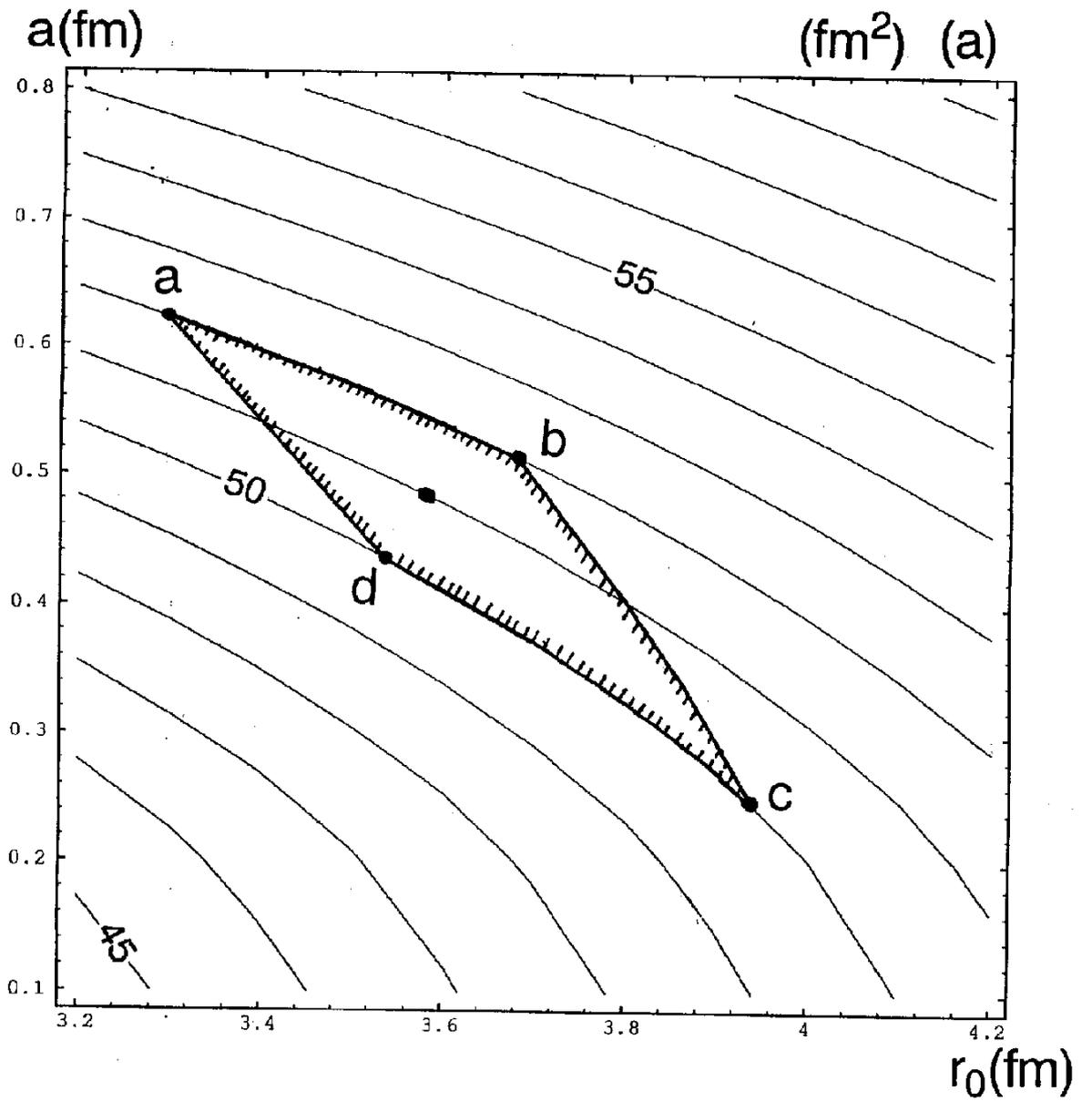
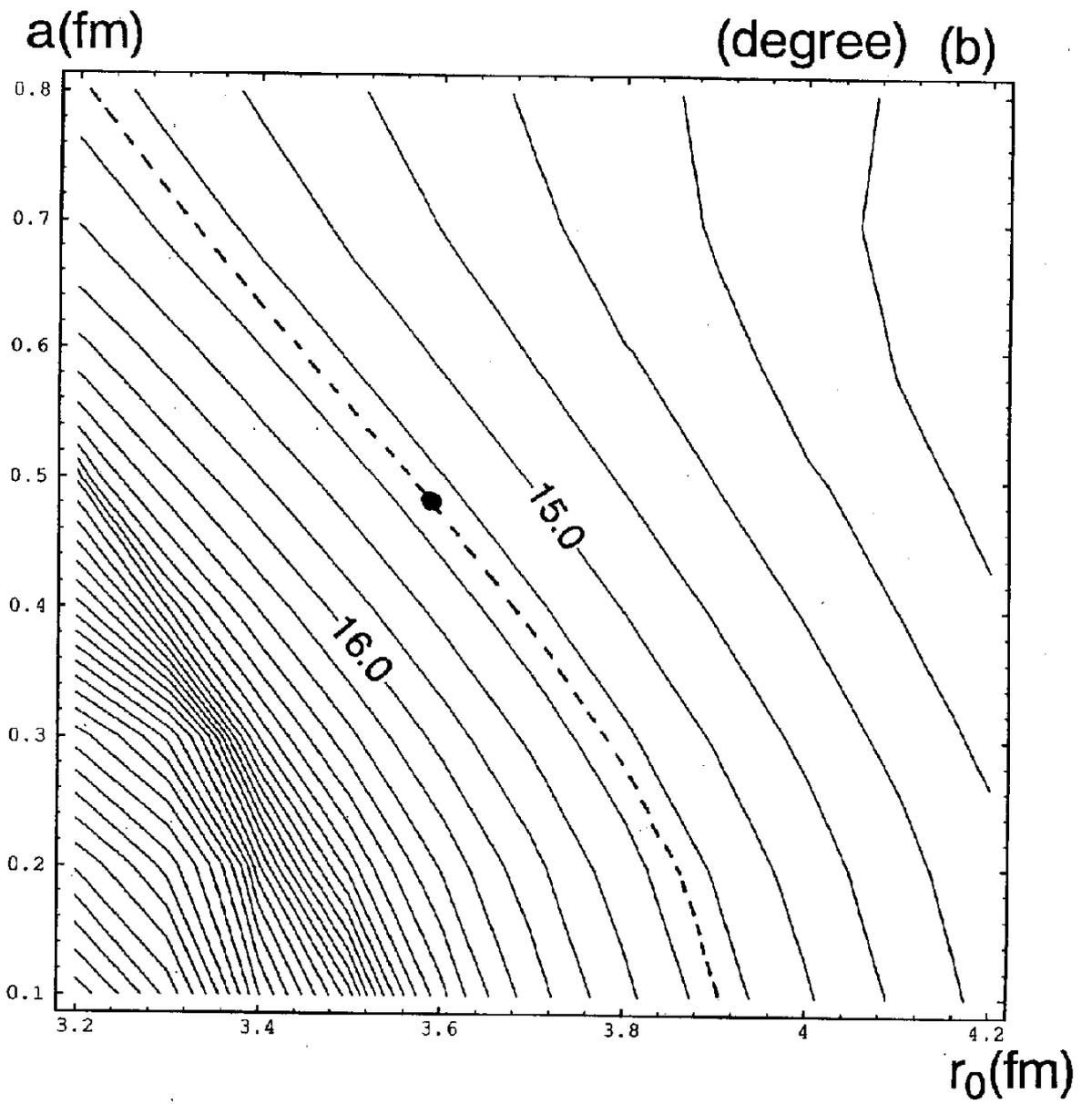
\item[Figure 4]
Contour plots of (a) reaction cross section (fm$^2$) and (b) first dip 
position of the differential cross section (degree) in the $r_0$ (radius 
parameter)-$a$ (diffuseness parameter) plane of Woods-Saxon neutron 
densities for 
$^{40}$Ca at T=300 MeV.  We use RMFT distribution \cite{13}
for the proton density.
Experimental values are shown by dotted lines in both plots, which
cross at the point indicated by the solid circle.  Experimental error is
assumed to be 2 \% and shown as the hatched area in (a).

\item[Figure 5]
Neutron density distributions of $^{40}$Ca determined from the reaction
cross sections and first dip angles are shown by solid lines.  Thick solid
line corresponds to the density which reproduce the both experimental values
shown as a solid circle in Fig.4.  Solid lines with a and c show the densities
correspond to points a and c in Fig.4 (a) .  For comparison,  the RMFT neutron
and proton density distributions are shown by dotted and dash-dotted
line, respectively.

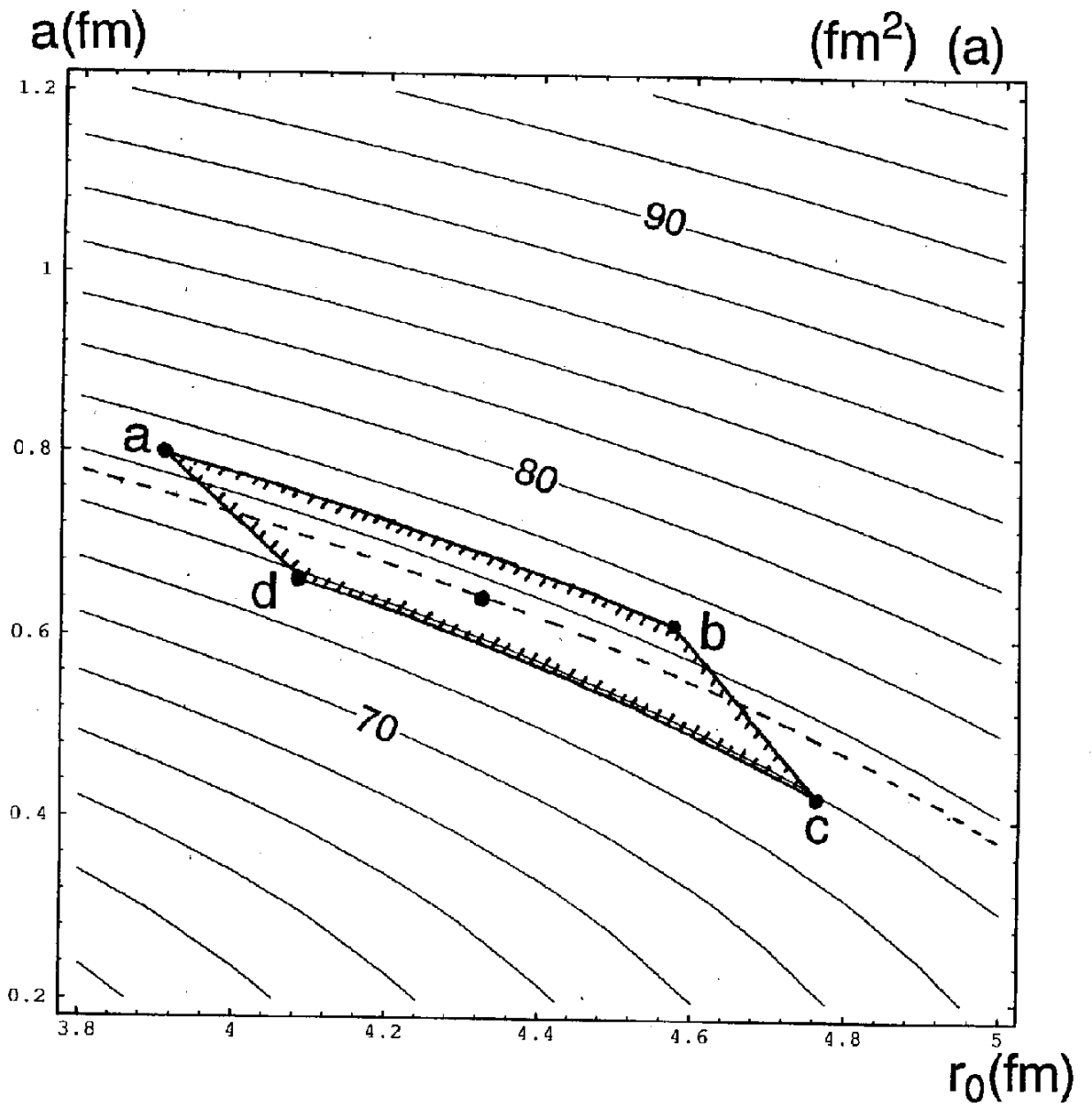
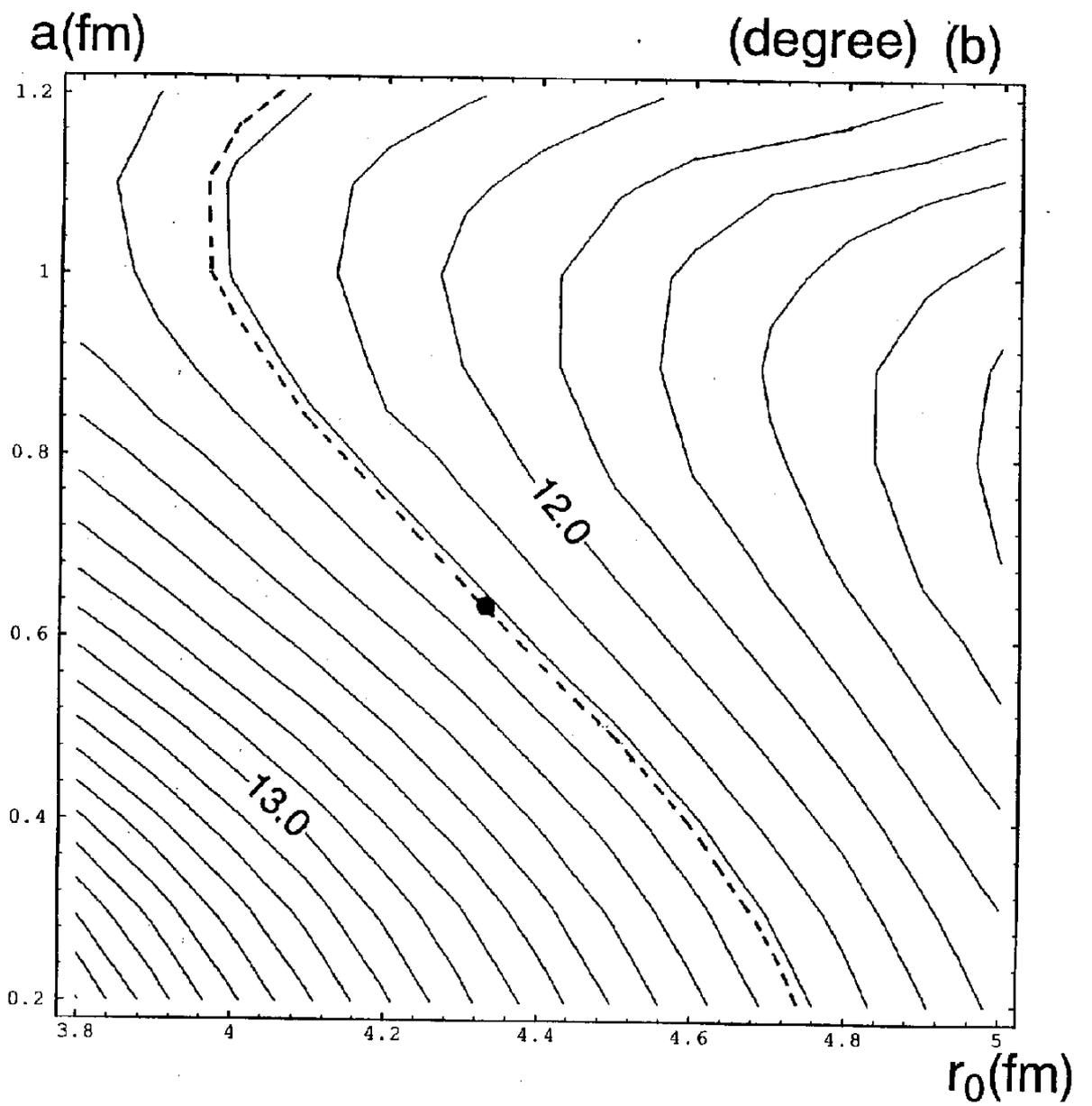
\item[Figure 6] 
Contour plots of (a) reaction cross section (fm$^2$) and (b) first dip 
position of the differential cross section (degree) in the $r_0$ (radius 
parameter)-$a$ (diffuseness parameter) plane of Woods-Saxon neutron densities for 
$^{60}$Ca at T=300 MeV.  We use RMFT distribution \cite{11} for the proton density.   
Dotted lines 
indicate the calculated values with RMFT neutron density, which cross at 
the point indicated by the black solid circle.  
The hatched area in (a) indicates 2 \% errors for the values with RMFT neutron density.

\item[Figure 7]
Neutron density distributions of $^{60}$Ca determined from the reaction
cross sections and first dip angles are shown by solid lines.  Thick solid
line corresponds to the density which reproduce the both values obtained
with the RMFT neutron density
shown as a solid circle in Fig.6.  Solid lines with a and c show the densities
correspond to points a and c in Fig.6 (a) .  For comparison,  the RMFT neutron
and proton density distributions are shown by dotted and dash-dotted
line, respectively.

\end{description}

\newpage
\centerline{Figure 1}

\newpage
\centerline{Figure 2}

\newpage
\centerline{Figure 3}

\newpage
\centerline{Figure 4 (a)}

\newpage
\centerline{Figure 4 (b)}

\newpage
\centerline{Figure 5}

\newpage
\centerline{Figure 6 (a)}

\newpage
\centerline{Figure 6 (b)}

\newpage
\centerline{Figure 7}

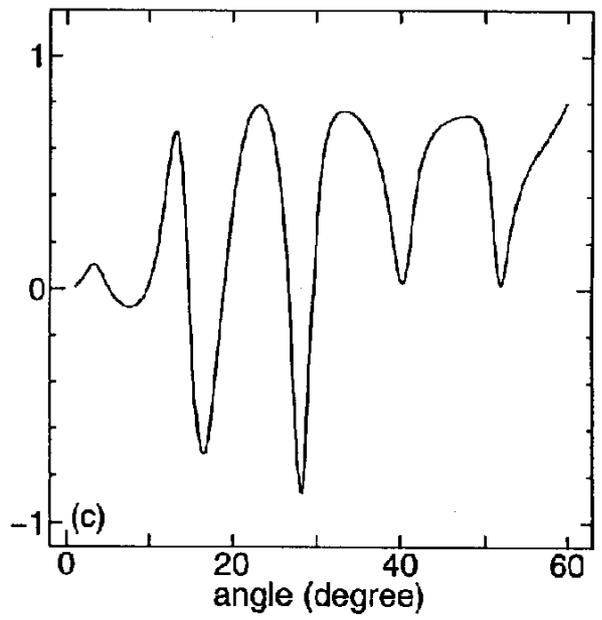
\end{document}